\documentclass[AMA,STIX1COL]{bioinfo}

\copyrightyear{2015} \pubyear{2015}

\usepackage{bm}          

\newcommand{\package}[1]{{\fontseries{b}\selectfont #1}}   

\access{Advance Access Publication Date: Day Month Year}
\appnotes{Manuscript Category}

\begin{document}
	\firstpage{1}
	
	\subtitle{Subject Section}
	
	\title[Omics Data Integration]{Bayesian Structural Equation Modeling in Multiple Omics Data Integration with Application to Circadian Genes}
	\author[Sample \textit{et~al}.]{Arnab Kumar Maity\,$^{\text{\sfb 1,}*}$, Sang Chan Lee\,$^{\text{\sfb 2}}$, Bani K. Mallick\,$^{\text{\sfb 2}}$, and Tapasree Roy Sarkar\,$^{\text{\sfb 2, 3}}$}
	\address{$^{\text{\sf 1}}$Early Clinical Development Oncology Statistics, Pfizer Inc., San Diego, 92121, USA, \\
		$^{\text{\sf 2}}$Department of Statistics, Texas A\&M University, College Station, 77843, USA, and \\
		$^{\text{\sf 3}}$Department of Biology, Texas A\&M University, College Station, 77843, USA.}
	
	\corresp{$^\ast$To whom correspondence should be addressed.}
	
	\history{Received on XXXXX; revised on XXXXX; accepted on XXXXX}
	
	\editor{Associate Editor: XXXXXXX}

\abstract{\textbf{Motivation:} It is well known that the integration among different data-sources is reliable because of its potential of unveiling new functionalities of the genomic expressions which might be dormant in a single source analysis. Moreover, different studies have justified the more powerful analyses of multi-platform data. Toward this, in this study, we consider the circadian genes' omics profile such as copy number changes and RNA sequence data along with their survival response. We develop a Bayesian structural equation modeling coupled with linear regressions and log normal accelerated failure time regression to integrate the information between these two platforms to predict the survival of the subjects. We place conjugate priors on the regression parameters and derive the Gibbs sampler using the conditional distributions of them.\\
	\textbf{Results:} Our extensive simulation study shows that the integrative model provides a better fit  to the data than its closest competitor. The analyses of glioblastoma cancer data and the breast cancer data from TCGA, the largest genomics and transcriptomics database, support our findings.\\
	\textbf{Keywords:} AFT Model, BRCA Tumor, Data Integration, GBM Tumor, TCGA. \\
	\textbf{Availability:} The developed method is wrapped in \textsf{R} package \package{semmcmc} available at \textsf{R} CRAN. \\
	A revised version is here: Maity, A. K., Lee, S. C., Mallick, B. K., \& Sarkar, T. R. (2020). Bayesian structural equation modeling in multiple omics data with application to circadian genes. Bioinformatics, 36(13), 3951-3958. https://doi.org/10.1093/bioinformatics/btaa286 \\
	\textbf{Contact:} \href{Arnab.Maity@pfizer.com}{Arnab.Maity@pfizer.com}\\
	\textbf{Supplementary information:} Supplementary data are available at \textit{Bioinformatics}
	online.}

\maketitle

\section{Introduction}

In the current era of precision medicine each subject is targeted for treatment modeled via individual healthcare data. To this end of advanced treatment it is of interest the molecular profiling besides the clinical profiling of the patients. Accurate prognostic prediction using molecular profiles is an essential ingredient to develop precision medicine. Under this regime, cancer studies that are focused on one-dimensional omics data have only provided limited information regarding the etiology of oncogenesis and tumor progression \citep{huang2017more}. To overcome this problem, scientists have focused to integrate multi-platform data in cancer research. 

The advent of multi-platform data has been directing the biological research and statistical methodological research to collect and analyze these multi-platform data. The Cancer Genome Atlas (TCGA) is the largest collection of parallel transcriptomics, genomics, and proteomics data along with patient's demographic information, primary aim of which is to generate, quality control, merge, analyze and interpret molecular profiles at the DNA, RNA, protein and epigenetic levels for hundreds of clinical tumors representing various tumor types and their subtypes \citep{weinstein2013cancer}. Cases that meet quality assurance specifications are characterized using technologies that assess the sequence of the exome, copy number variation (CNV, measured by SNP arrays), DNA methylation, mRNA expression and RNA sequence, microRNA expression and transcript splice variation, whole genome sequencing and reverse phase protein arrays (RPPA). Attention is being paid to identify the genomic alterations across these platforms to improve the therapeutic response which may be evident from the phenotypical measures such as survival of the cancer patients. The reasoning behind this attention can be motivated by each of the hundreds of genetic alterations inside of a genome providing a complementary view of the underlying complex biological process and thus an integrative analysis of multiple platform is required to achieve the overreaching goal of cancer studies.

Circadian oscillation is a fundamental process that regulates a wide variety of physiological and metabolic processes.  Perturbations of circadian rhythmicity is associated with significant physiological consequences including metabolic disorders and cancer \citep{sahar2009metabolism}. Increased cancer incidence and progression have often been linked to disruption or deregulation of the molecular mechanism of the circadian clock \citep{fu2013circadian}. Circadian rhythms are referred to those organisms which exhibit time dependent behavior across a 24 hour day. These outputs are driven by manifestations of phasic cyclic gene expression patterns. Nearly half of all protein-coding genes show circadian-dependent transcription in at least one tissue in mammals \citep{andreani2015genetics}. There is increasing evidence that links dysfunction of the clockwork with the pathogenesis of cancer such as breast cancer and brain cancer \citep{davis2006circadian}. In this article, we propose a Bayesian structural equation coupled with Bayesian accelerated failure time (AFT) model to carry out an integrative analysis where the integration takes place among the multiple platform of omics data. We consider some important circadian genes which have been reported to play an important role in breast and brain cancer progression. 

We note that the direction of biological relationship is arbitrary and it may be a good practice to introduce some latent variables along with the observed variables to describe the relationships. To this end, \cite{wong2018efficient} proposed structural equation modeling (SEM) to model the TCGA data. The history of SEM dates back to \cite{bentler1980linear} and it has been used extensively in the literature thereafter; for example, in psychology \citep{quintana1999implications}, in economics \cite{heckman2005structural}, in healthcare sector \cite{naliboff2012gastrointestinal}. Structural equation modeling requires the introduction of latent variables and there are several studies which make use of latent variable for survival regressions, for example in cox proportional hazard model \citep{stoolmiller2006modeling, larsen2005cox}.

The concept of integration is very broad. Generally, based on the direction, they can be classified into two broad groups -- horizontal and vertical \citep{chu2017integrated}. In the horizontal integration analysis omics data of same types but different studies or laboratories are combined. On the other hand, when the different omics data for the same patient is analyzed then it is called the vertical integration, which is the focus of our current study. The vertical integration methods are then categorized into different groups depending on the methodologies used. For instance, Bayesian and non-Bayesian integration methods, network based integration method, supervised learning and non-supervised learning etc. For a full review we refer the readers to \cite{huang2017more}. Other comprehensive references include \cite{tseng2015integrating, gomez2014data, hamid2009data}. A popular network based method was reported by \cite{vaske2010inference}, who developed a supervised graphical model incorporating the pathway information. Another example of unsupervised learning is iCluster method by \cite{shen2009integrative}, where by using the penalized likelihood approach they derived a clustering solution for tumor cells. \cite{daemen2009kernel} proposed a kernel based support vector machine to integrate microarray and omics data for the cancer patients. However, many of these methods do not consider the underlying biological relationship between multiple omics data sources.

As a remedy, \cite{wang2012ibag} proposed an integrated Bayesian model which essentially combines a two stage regressions in a unified manner. The first model regresses the gene expression on the methylation expression, and the second model then regresses the clinical variable or the phenotype on the estimated effects from the first model. However, a major criticism of this model is that statistically it encourages increment of the errors when going from the first model to the second model.  

Nevertheless, the Bayesian paradigm for structural equation is notably scant;  important references include \cite{palomo2007bayesian, song2012tutorial}. Among them the work of Song and Lee \cite{song2012tutorial} have described the basic ingredients of Bayesian structural equation modeling with few examples and the codes are written in WinBUGS. However, to the best of our knowledge, there has been no study on the application of structural equation modeling under the Bayesian regime in survival settings. In this article, we propose a Bayesian structural equation coupled with Bayesian accelerated failure time (AFT) model to carry out an integrative analysis where the integration takes place among the multiple platform of omics data. We consider the DNA copy number variation and RNAseq data-sources as the two platforms to predict the survival of the patients. We show that an integrative analysis outperforms the usual regression model where the underlying biological relationship is not captured. 

\begin{figure}[h]
	\centering
	\includegraphics[height = 6cm, width = 8 cm]{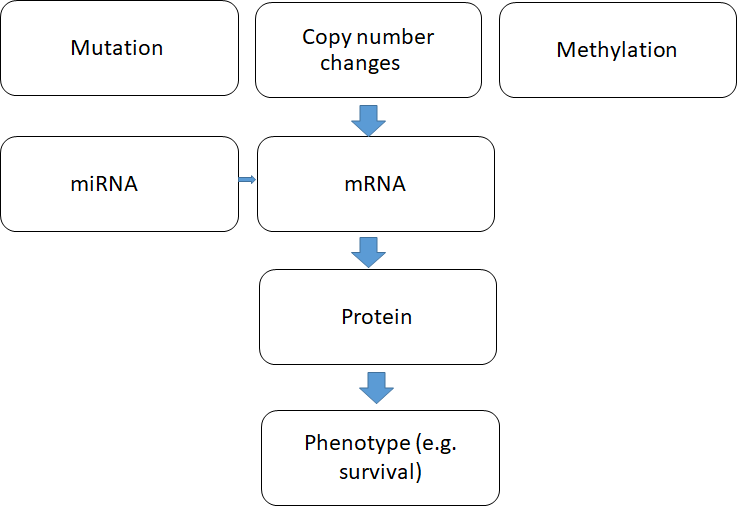}
	\caption{Biological relationships among gene expressions data platforms.}
	\label{figure_flow}
\end{figure}

In general, the TCGA collects and provides various levels DNA level data -- methylation expression, mutation, and DNA copy number changes \cite{wang2012ibag}. These molecular features coupled with microRNA (miRNA) expressions data are known to affect the gene expression level data measured by microarray technology or by next generation RNA sequencing technology. The genes then code for proteins which directly controls the tumor growth. This relationship is schematically displayed in Diagram \ref{figure_flow}. Any integration method which wants to consider the underlying direction among the platforms faces additional challenge and thus requires additional processing. For example, each transcriptomics factor may or may not be responsible for over-expression or under-expression for either one, or multiple or neither of the genomics features. In a very similar fashion, each gene expression may or may not code proteins and can affect the function of multiple proteins which are the primary factors for tumor growth or tumor suppression. To overcome this, we assume that the expressions of each platform is controlled by a latent variable and as well as the latent variables from other platforms. The details are described in Section \ref{section_model}. 

In this study, we introduced the Bayesian methodologies of TCGA data using a structural equation model and used the posterior analysis via the Markov Chain Monte Carlo (MCMC). Our model formulation is similar in the spirit of what is proposed by \cite{wong2018efficient}. However, they considered Cox proportional hazard model in order to model the survival time and used the EM algorithm technique to maximize the likelihood function. In addition, they assumed the latent variable can be measured via the various types of gene expressions for a single gene and hence their model could consider a single gene at a time. In the contrary, we assume that for multiple genes there exist a latent variable for circadian gene expressions or copy number variations, so this can easily accommodate multiple genes in a single model for a better result. We considered Glioblastoma cancer and Breast cancer data sets for a set of genes which have been known to affect the circadian rhythms. For integration among different platforms we consider two types of measurements of those genes -- copy number changes and normalized RNA sequencing data. Our model showed that integration among these two platforms provide a better fit for the survival outcome of the subjects. 

The remainder of this article is organized as follows. Section \ref{section_model} introduces the Bayesian methodologies of TCGA data using a structural equation model and provides a brief description how to carry the posterior analysis via the Markov Chain Monte Carlo (MCMC). In Section \ref{section_simulation} we describe simulation examples and show that the performance of our proposed model is superior than the general kind of regression. We then illustrate our methodology by applying to TCGA cancer data in Section \ref{section_real_data}. We consider two cancers namely Glioblastoma cancer and Breast cancer data sets for a set of genes who have been known to affect the circadian rhythms of a human biological clock. For integration among different platforms we consider two types of measurements of those genes -- copy number changes and normalized RNA sequencing data. Using our method we justify that integration among these two platforms provide a better fit for the survival outcome of the subjects. The discussions and conclusions in Section \ref{section_conclusion} are then followed.

\section{Multiple Omics Data Integrated Model}  \label{section_model}

\subsection{The Model} 

Vertical integration is referred to the analysis when the different data are collected from multiple transcriptomics, genomics, and proteomics platforms for a same subject to infer about the cell outcomes. To ease of explanation, we provide the model development strategy for two platforms namely copy number variation (CNV) and gene expression (mRNA), however can be generalized for multiple platforms in a straightforward manner. The phenotypical model we consider here is the log normal AFT model for the survival outcome with demographic variables as the covariates. 

In what follows, we assume that each platform gets affected and can be explained by a latent variable. Let $ n $ be the number of individuals, $ q_1 $ be the number of mRNA expressions, and $ q_2 $ be the number of CNV measurements. Also let $ \eta_1 $ and $ \eta_2 $ be the latent variables which control the mRNA expressions, $ \bm{U}_{1k}, k = 1, \ldots, q_1 $, and CNV measurements, $ \bm{U}_{2l}, l = 1, \ldots, q_2 $, respectively. Each gene measurement is for $ n $ individuals and hence a $ n \times 1 $ vector. In addition, $ \eta_1 $ can also be explained by $ \eta_2 $, meaning that the significance of the copy number changes are captured to describe the mRNA expressions. Finally, we construct the AFT regression model of survival data $ (\bm{t}^*, \bm{\delta}) = ((t^*_1, \delta_1), \ldots, (t^*_n, \delta_n))' $ with some covariates $ \bm{x}_j = (x_{1j}, \ldots, x_{nj})', j = 1, \ldots, p $. 

\begin{figure}[h]
	\centering
	\includegraphics[height = 3cm, width = 8 cm]{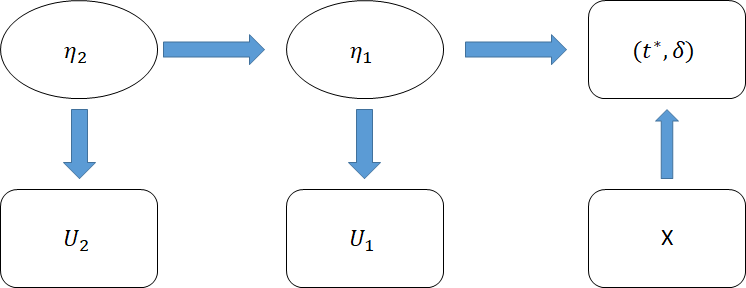}
	\caption{Biological relationships among gene expressions data platforms.}
	\label{figure_sem}
\end{figure}

Here $ \delta_i $ is the censored indicator and takes 1 if a death is observed and takes 0 if right censored. Given the actual death time $ \bm{t} $  and censoring time $ \bm{c} $ are independent, $ t^*_i = \min(t_i, c_i) $. Hence, we propose the following structural equation model:
\begin{align}
	\log \bm{t} =& \alpha_t + \bm{X} \bm{\beta}_t + \eta_1 \phi_t + \bm{\epsilon}_t  \nonumber \\
	\bm{U}_{1k} =& \alpha_{u_1k} + \eta_1 \phi_{u_1k} + \bm{\epsilon}_{u_1k}, \quad k = 1, \ldots, q_1   \label{equation_mRNA} \\
	\bm{U}_{2l} =& \alpha_{u_2l} + \eta_2 \phi_{u_2l} + \bm{\epsilon}_{u_2l}, \quad l = 1, \ldots, q_2   \label{equation_CNV}.
\end{align} 
Here, $ \epsilon_t $ is the error vector for the AFT regression model. Assuming $ \bm{\epsilon}_t \sim N(0, \sigma_t^2 \bm{I}) $ gives raise to the log normal AFT model. In addition, we assume $ E(\bm{\epsilon}_{u_1}) = E(\bm{\epsilon}_{u_1}) = 0 $ and $ Cov(\bm{\epsilon}_{u_1}, \bm{\epsilon}_{u_2}) = 0 $. $(\alpha_t, \bm{\alpha}_{u_1}, \bm{\alpha}_{u_2}) $ are the intercept parameters and $ (\bm{\beta}_t, \phi_t, \bm{\phi}_{u_1}, \bm{\phi}_{u_2}) $ are the suitable regression parameters. To carry out the analysis in Bayesian fashion we impute the censored observations from the appropriate truncated normal distribution. In addition, we assume that $ \bm{\epsilon}_{u_1k} \sim N(0, \sigma_{u_{1k}}^2I) $ and $ \bm{\epsilon}_{u_2l} \sim N(0, \sigma_{u_{2l}}^2I) $ such that each of (\ref{equation_mRNA}) and (\ref{equation_CNV}) is a standard linear regression model. Furthermore, while it is assumed that $ \eta_1 $ is dependent on $ \eta_2 $ via $ \eta_1 \sim N(\eta_2, \sigma_{\eta_1}^2) $, $ \eta_2 $ assumed to be independently follow $ N(0, \sigma_{\eta_2}^2) $. The schematic diagram of our structural equation model is shown in Figure \ref{figure_sem}. 

We assume the standard multivariate normal distribution on the regression coefficients $ \bm{\beta} $ which can be made a vaguely informative by assuming a large variance component in the variance covariance matrix. Nevertheless, other regression parameters $ \alpha_t, \alpha_{u_1k}, \alpha_{u_2l}, \phi_{u_1k}, \phi_{u_2l} $ are all assumed to follow a normal distribution and can be made a vaguely informative. While we assume a noninformative prior on $ \sigma_t^2 $, the other variance parameters are kept as fixed for our study. With these ingredients, the full Bayesian hierarchical representation is
\begin{align}
	\log \bm{t} \sim & N(\alpha_t + \underline{\bm{X}} \bm{\beta}_t + \eta_1 \phi_t, \sigma_t^2 \bm{I}) \label{equation_time_distribution}  \\
	\bm{U}_{1k} \sim & N(\alpha_{u_{1k}} + \eta_1 \phi_{u_{1k}}, \sigma_{u_1}^2 \bm{I}), \quad k = 1, \ldots, q_1 \label{equation_mRNA_distribution}  \\
	\bm{U}_{2l} \sim & N(\alpha_{u_{2l}} + \eta_2 \phi_{u_{2l}}, \sigma_{u_2}^2 \bm{I}), \quad l = 1, \ldots, q_2  \label{equation_CNV_distribution}  \\
	\eta_1 \sim & N(\eta_2, \sigma_{\eta_1}^2)  \label{equation_eta1_distribution}   \\
	\eta_2 \sim & N(0, \sigma_{\eta_2}^2)  \label{equation_eta2_distribution}  \\
	\bm{\beta} \sim & N(\bm{\beta}_{t0}, \sigma_t^2 \Sigma_\beta)   \nonumber  \\
	\alpha_t \sim & N(\alpha_{t0}, \sigma_t^2 \sigma_{\alpha_t}^2)  \nonumber  \\
	\alpha_{u_{1k}} \sim & N(\alpha_{u_{1k}0}, \sigma_{\alpha_{u_{1k}0}}^2), \quad k = 1, \ldots, q_1  \nonumber  \\
	\alpha_{u_{2l}} \sim & N(\alpha_{u_{2l}0}, \sigma_{\alpha_{u_{2l}0}}^2), \quad l = 1, \ldots, q_2  \nonumber  \\
	\phi_t \sim & N(\phi_{t0}, \sigma_\phi^2)  \nonumber  \\
	\phi_{u_{1k}} \sim & N(\phi_{u_{1k}0}, \sigma_{\phi_{u_{1k}0}}^2), \quad k = 1, \ldots, q_1  \nonumber  \\  
	\phi_{u_{2l}} \sim & N(\phi_{u_{2l}0}, \sigma_{\phi_{u_{2l}0}}^2), \quad l = 1, \ldots, q_2  \nonumber  \\  
	\sigma_t^2 \sim & \pi(\sigma_t) \equiv 1/\sigma_t^2.  \nonumber  
\end{align}

\subsection{Identifiability}

A common problem is identifiability of the full model when using a structural equation models. \cite{bollen2009two} discussed few conditions under which a structural equation model becomes identifiable using the \textit{exogenous X} rule. In what follows, we provide a brief descriptions of the conditions and show that those hold under our formulation of the model.

First, each latent variable should have at least one observed variable that loads solely on it and the associated errors of measurement are uncorrelated. According to the formulation of our model the observed matrices $ \underline{\bm{U}_1} $ and $ \underline{\bm{U}_2} $ are solely related to the latent variables $ \eta_1 $ and $ \eta_2 $ respectively. In addition, we have assumed that the corresponding error vectors are uncorrelated. So this suffices this condition. Second, each latent variable must have at least two observed indicators in total and the errors of these other indicators are uncorrelated with those of the unique indicators. This is satisfied trivially with the formulation of our model. Finally, the latent variable model (\ref{equation_eta1_distribution}) must have an identical structure which is also true here.

\subsection{Posterior Computation}  \label{section_posterior}

The posterior computation for the right censored data is not straightforward since the censored data are not originally observed. Nonetheless, we impute the right censored observations following the data augmentation approach \citep{bonato2010bayesian, tanner1987calculation}. We denote the augmented data vector by $ \bm{y} = (y_1, \ldots, y_n)' $ where
\[
\left\{
\begin{array}{l l}
y_{i} = \log t_{i}^* \quad \text{if } \delta_{i} = 1 \\
y_{i} > \log t_{i}^* \quad \text{if } \delta_{i} = 0. \\
\end{array} \right.
\]
Hence, to carry out the posterior analysis, at the $s$-th iteration of the MCMC chain, the censored data are sampled from the truncated normal distribution
\[
y_i^{(s)} \sim N(\alpha_t^{(s)} + \sum_{j = 1}^p x_{ij} \beta_{tj}^{(s)}, {\sigma_t^2}^{(s)}) I(y _i > \log t_i^*)   \text{ if } \delta_i = 0.
\]

In a similar fashion, with the latent variables in the joint likelihood the posterior distribution becomes intractable. However, using the data augmentation scheme, the latent variables can be updated from the conditional distributions $ \eta_1|. \sim N({\mu_{\eta_1}}_{\text{post}}, {\sigma_{\eta_1}}_{\text{post}}^2) $ and $ \eta_2|. \sim N({\mu_{\eta_2}}_{\text{post}}, {\sigma_{\eta_2}}_{\text{post}}^2) $ respectively, where,
\begin{align*}
	{\sigma_{\eta_1}}_{\text{post}}^2 =& 1/\sigma_{\eta_1}^2 + (\sum_{k = 1}^{q_1} \phi_{u_{1k}}^2/\sigma_{u_{1k}}^2) \\
	{\mu_{\eta_1}}_{\text{post}} =& 1/{\sigma_{\eta_1}}_{\text{post}}^2 (\eta_2/\sigma_{\eta_1}^2 + (\sum_{k = 1}^{q_1} \phi_{u_{1k}} \bm{u}_{1k}/\sigma_{u_{1k}}^2) + \\
	& \phi_t \bm{1}_n^T \log \bm{t}/\sigma_t^2 - \sum_{k = 1}^{q_1} (\alpha_{u_{1k}} \phi_{u_{1k}} /\sigma_{u_{1k}}^2) -   \\
	& \phi_t \alpha_t/\sigma_t^2 - \bm{\beta}_t^T \underline{\bm{X}}^T \phi_t \bm{1}_n /\sigma_t^2) \\
	{\sigma_{\eta_2}}_{\text{post}}^2 =& 1/\sigma_{\eta_1}^2 + 1/\sigma_{\eta_2}^2 + \sum_{l = 1}^{q_2} (\phi_{u_{2k}}^2/\sigma_{u_{2k}}^2)  \\
	{\mu_{\eta_2}}_{\text{post}} =&  1/{\sigma_{\eta_2}}_{\text{post}}^2 (\eta_1/\sigma_{\eta_1}^2 + \sum_{l = 1}^{q_2} (\phi_{u_{2l}} u_{2l}/\sigma_{u_2l}^2) +  \\
	& \sum_{l = 1}^{q_2}(\alpha_{u_2l} \bm{\phi}_{u_2l}/\sigma_{u_{2l}}^2). \\ 
\end{align*}

Once the latent variables are updated they can be treated as observed as the other observed variables. Hence, the conditional distributions are available explicitly, and Gibbs sampling can be employed to cycle the iterations. The remaining details are as follows. 

When, $ \bm{\beta_{t0}} = \alpha_{t0} = \phi_{t0} = 0 $,
\begin{align*}
	\bm{\beta}_t|. \sim & N(B^{-1} \underline{\bm{X}}^T (\bm{y} - \alpha_t - \eta_1 \phi_t), \sigma_t^2 B^{-1}) , \quad B = (\underline{\bm{X}}^T \underline{\bm{X}} + \Sigma_\beta^{-1})  \\
	\alpha_t|. \sim & N(A^{-1} \bm{1}_n^T (\bm{y} - \underline{\bm{X}} \bm{\beta}_t - \eta_1 \phi_t), \sigma_t^2 A^{-1}), \quad A = (\bm{1}^T \bm{1} + \sigma_{\alpha_t}^2)  \\
	\phi_t|. \sim & N(P^{-1} \eta_1 \bm{1}_n^T (\bm{y} - \alpha - \underline{\bm{X}} \bm{\beta}_t), P^{-1}), \quad P = (\eta_1 \bm{1}^T \bm{1} + \sigma_{\phi_t}^2)   \\
	\sigma_t^2 |. \sim & \text{Inverse Gamma} [\text{shape} = (n + p + 1)/2,   \\
	& \text{scale} = \{(\bm{y} - \alpha_t - \underline{\bm{X}} \bm{\beta}_t - \eta_1 \phi_t)^T (\bm{y} - \alpha_t - \underline{\bm{X}} \bm{\beta}_t - \eta_1 \phi_t)  \\
	&  + \bm{\beta}_t^T \bm{\beta}_t + \alpha_t^2 \}/2]
\end{align*}

Other parameters can be can be updated from similar conditional distributions by assuming $ \eta_1 $ and $ \eta_2 $ as observed once they are imputed in the MCMC chain.

\subsection{Goodness of Fit}

There exist several model validation criteria such as log pseudo marginal likelihood, \citep{gelfand1992model}, L-measure \citep{ibrahim1994predictive}, or DIC \citep{spiegelhalter2002bayesian}. The literature is also advocated with the application of these criteria in survival settings, for example, see \cite{brown2005flexible, ibrahim2005bayesian, rizopoulos2011bayesian}. In this paper, to measure the goodness of fit, we consider the Deviance Information Criterion which combines goodness of fit of a model with a penalty for model complexity and is defined, as the model deviance + $2 \times$ (effective number of parameters), evaluated at a posterior point estimate of the parameter. In particular, $ \hbox{$ \text{DIC}=D(\overline{\theta})+2p_{\text{D}}, $} $ where  $ \hbox{$ D(\theta)=-2 \log f(.|\theta)$, $f(.|\theta) $} $ is the likelihood function of the  model and $\overline{\theta}$ is an estimate of the model parameter $\theta$. In the above expression $p_{\text{D}}$ is termed as the effective number of parameters and is defined as $ \hbox{$ p_{\text{D}}=\overline{\text{D}(\theta)}-D(\overline{\theta}) $} $, where $ \hbox{$ \overline{\text{D}(\theta)} $} $ is a posterior point estimate of the deviance. In our proposed model it is possible to partition the likelihoods over survival and coordinates and thus to obtain the DIC for survival model. A model with smaller value of DIC is preferred. 

The conditional predictive ordinate (CPO) is a Bayesian model diagnostic criterion introduced in \cite{geisser1979predictive} and its implementation in sampling based approaches is discussed in \cite{gelfand1992model}. For a model, the conditional predictive ordinate of the $i^{th}$ observation $y_i$ is defined as
\[
\text{CPO}_{i} =f(y_{i}|\bm{y}_{-i})=\int f(y_{i}|\bm{\theta})\pi(\bm{\theta}|\bm{y}_{-i})d\bm{\theta},
\]
where $\bm{y}_{-i} = \bm{y} \setminus \{y_i\}$. \cite{gelfand1992model} provided an estimate of CPO$_i$ based on Markov chain samples from the full posterior $\pi(\bm{\theta}|\bm{y})$. The Log Pseudo Marginal Likelihood (LPML) of model $ \mathcal{M}(\alpha)$ is constructed similar to the log-likelihood, but based on the CPO$_i$, and is defined as
$\text{LPML} = \log \prod _{i=1}^{n} \text{CPO}_i$. Model with higher LPML is preferred. The LPML is well defined provided the predictive density is proper and thus may be defined under improper priors as well.

\section{Operating Characteristics in Simulation Studies} \label{section_simulation}

\subsection{Integrated Model as the Data-Generating Model}  \label{section_simulation_integrated}
In this section we study some simulated examples to observe the prediction performance using our Bayesian structural equation integrated model. To this direction we generate the covariate matrix from a multivariate normal distribution with mean $ \bm{0} $, variance covariance matrix as unit matrix, and dimension 2, i.e., p = 5. All the regression parameters $ \bm{\beta} $ and the intercept parameters $ \alpha_t, \bm{\alpha}_{u_{1}}, \bm{\alpha}_{u_{2}} $ are generated from a Uniform distribution $ U(-1, 1) $. We set the latent variable coefficients $ \phi_t = \phi_{u_1} = \phi_{u_2} = 1 $ and the variance parameters $ \sigma_t^2 = 1 $, $ \bm{\sigma}_{u_1}^2 = \bm{\sigma}_{u_2}^2 = \bm{1} $. Additionally, we consider generating the data by setting $ \sigma^2 = 2 $ and $ \bm{\sigma}_{u_1}^2 = \bm{\sigma}_{u_2}^2 = 1 $. The case of varying $ \bm{\sigma}_{u_1}^2 $ and $ \bm{\sigma}_{u_2}^2 $ is discussed in Section \ref{section_varying_variance}, and the impact of placing a informative proper prior is discussed in the supplementary material. 

Then the latent variables $ \eta_1 $ and $ \eta_2 $ are generated according to (\ref{equation_eta1_distribution}) and (\ref{equation_eta2_distribution}) respectively. The mRNA expressions $ \underline{\bm{U}}_1 $ and copy number changes $ \underline{\bm{U}}_2 $ are simulated according to (\ref{equation_mRNA_distribution}) and (\ref{equation_CNV_distribution}) respectively. Finally we simulate the time components $ \bm{t} $ in log scale according to (\ref{equation_time_distribution}). We consider both situations with censoring and with no censoring. When censored subjects are created the censoring distribution of the censoring time $ \bm{c} $ is assumed to follow a Gamma distribution and hence the amount of censored data can be controlled by varying the shape and scale parameters of the Gamma distribution. So, we obtain the observed paired response data $ \{t_i^*, \delta_i\} = \{\min(t_i, c_i), I(t_i < c_i) \}, i = 1, \ldots, n $, where, $ n = 100 $. We simulate 100 similar datasets in order to assess the goodness of fit of the integrated model in repeated experiments. 

When fitting the integrated model to the simulated data we set all the mean parameters of the prior distributions as $ 0 $. In addition, the variance parameters of the normal priors are kept as 1 while the variance covariance matrix for $ \bm{\beta}_t $ is $ \text{diag}(100, 100) $. We simulate the datasets with the censoring rates 0\% (no censoring), 28\%, 37\%, and 50\%. It is observed that in the Bayesian analysis after discarding $ 2000 $ burn-in samples $ 100000 $ iterations with $ 100 $ thinning provides a good stationary Markov chain. For comparison, we also fit a Bayesian log normal AFT model on the data with covariates $ \underline{\bm{X}}, \underline{\bm{U}}_1, $ and $ \underline{\bm{U}}_2 $, i.e. the demographic variables, mRNA, and CNV data respectively and referred to it as nonIntegrated model. In particular, we fit the model 
\begin{equation}  \label{equation_nonIntegrated_model}
\log \bm{t} = \alpha + \underline{\bm{X}} \bm{\beta} + \underline{\bm{U}}_1 \bm{\gamma}_1 + \underline{\bm{U}}_2 \bm{\gamma}_2 + \bm{\epsilon},
\end{equation} 
where $ \bm{\epsilon} \sim N(0, \sigma^2 I) $, and $ \bm{\beta}, \bm{\gamma}_1, \bm{\gamma}_2 $ are corresponding regression coefficients. We impose vaguely informative prior on the parameters as discussed previously and to carry out the Bayesian analysis we augment the censored data and impute them.

To the best of our knowledge the existing software do not handle censored survival outcomes. Nevertheless, to compare with the available software packages we selected the \textsf{R} package \package{lavaan} \citep{Rosseel2012lavaan} as a representative. This method is used to estimate the MSE when there is no censoring in the data. A related comparison with the iBAG method \citep{wang2012ibag} is provided in the supplementary material.

\begin{table}[h]
	\centering
	\caption{Goodness of fit for the integrated and nonIntegrated models in simulation examples.}
	\begin{tabular}{llccc}
		\hline
		\hline
		Censor Rate & Method & DIC & LPML & MSE \\
		\hline
		\hline
		\multicolumn{5}{c}{$ \sigma_t^2 = 1 $} \\
		\hline
		0\%  & integrated    & 276.32 & -575.22 & 0.000 \\
		     & nonIntegrated & 284.42 & -802.14 & 0.000 \\
		     & lavaan        & --     & --      & 2.362 \\
		     \hline
		28\% & integrated    & 290.04 & -313.82 & 0.048  \\
		     & nonIntegrated & 314.25 & -787.30 & 0.052\\
		\hline
		37\% & integrated    & 288.07 & -283.90 & 0.113 \\
		     & nonIntegrated & 308.30 & -815.26 & 0.221  \\
		\hline
		50\% & integrated    & 300.14 & -235.73  & 0.296 \\
		     & nonIntegrated & 303.02 & -1267.44 & 0.345 \\
		\hline
		\multicolumn{5}{c}{$ \sigma_t^2 = 2 $} \\
		\hline
		0\%  & integrated    & 417.46 & -344.67 & 0.000 \\
		     & nonIntegrated & 441.50 & -972.82 & 0.000 \\
		     & lavaan        & --     & --      & 4.154 \\
		\hline
		28\% & integrated    & 395.24 & -255.26 & 0.623  \\
		     & nonIntegrated & 409.13 & -556.66 & 0.734 \\
		\hline
		37\% & integrated    & 386.99 & -197.14 & 0.946 \\
		     & nonIntegrated & 397.52 & -467.39 & 1.095  \\
		\hline
		50\% & integrated    & 356.88 & -159.57 & 1.345 \\
		     & nonIntegrated & 376.53 & -491.02 & 1.480 \\
		\hline
		\multicolumn{5}{c}{Data-generating model is the nonIntegrated Model} \\
		\hline
		0\%  & integrated    & 553.18 & -497.72 & 0.000 \\
		     & nonIntegrated & 545.42 & -441.32 & 0.000 \\
		     & lavaan        & --     & --      & 5.284 \\
		\hline
		28\% & integrated    & 508.00 & -285.73 & 1.834  \\
	 	     & nonIntegrated & 503.68 & -257.19 & 1.787 \\
		\hline
		37\% & integrated    & 499.62 & -246.98 & 2.567 \\
		     & nonIntegrated & 495.85 & -236.26 & 2.551 \\
		\hline
		50\% & integrated    & 486.30 & -240.78 & 4.088 \\
		     & nonIntegrated & 483.29 & -209.78 & 4.031 \\
		\hline
		\hline
	\end{tabular}
	\label{table_simulation}
\end{table}

The MCMC routine takes about 1.6 minutes per 100000 iterations in a computing system equipped with Intel(R) Core(TM) i5-8350U CPU @ 1.76 GHz 1.90 GHz processor, 8.00 GB RAM, and 64-bit operating system. Table \ref{table_simulation} summarizes the result and the superior performance of the proposed integrated model is evident form the Table. For instance, in the case of $ \sigma_t^2 = 1 $, when about 28\% data is right censored the DIC of integrated model is 290.04 while the same for the nonIntegrated model is 314.25; this suggests that the integrated method where the underlying relationship is captured, provides a better fit to the data. Similarly, the LPML due to integrated method is -313.82 that is greater than -787.30, LPML due to the nonIntegrated method, which supports in favor of the structural equation model based integration method. 

Furthermore, the existing \package{lavaan} package employs a non Bayesian method to fit the SEM and hence the model fitting criteria such as DIC and LPML can not be computed. When the MSE is calculated for the non censoring case we notice that the estimated average MSE 2.36 is far larger than that of integrated model. Moreover, for all the censoring cases and non censoring cases the MSE due to the integrated model remains smaller than the nonIntegrated model.

\subsection{NonIntegrated Model as the Data-Generating Model}  

This section is devoted to the simulation study when the data is generated from the nonIntegrated model (\ref{equation_nonIntegrated_model}). We use this model to generate 100 simulated datasets in which the data generation scheme was very similar to what have been discussed in Section \ref{section_simulation_integrated}. After generating the data set according to a nonIntegrated model we fit both integrated model and nonIntegrated model. We provide the summary of the results in Table \ref{table_simulation}. We note that, for instance, when there are about 28\% of the data is right censored and a nonIntegrated model is fitted in the generated datasets the average DIC is 503.68 and when the integrated model is fitted then the average DIC is 508.00. Hence, it can be concluded that even though the integrated model does not provide a better fit the difference is, however, very small to distinguish unlike the case when the data generating model is the integrated model. This phenomena is evident in the other results of DIC, LPML, and MSE in Table \ref{table_simulation}.  

\subsection{Sensitivity Analysis}  \label{section_varying_variance}

The purpose of this example to examine the effect in the performance of our proposed integrated model under different fixed values of the variance parameters $ \bm{\sigma}_{u_1} $ and $ \bm{\sigma}_{u_2} $. We generate the data in the same way as in Section \ref{section_simulation_integrated}. The censoring distribution parameters are set in such a way that the average censoring for 100 simulated data is about 25\%. The other priors were similar to what we had in the previous section. Table \ref{table_sensitivity} presents the results under different set of values of  $ \bm{\sigma}_{u_1} $ and $ \bm{\sigma}_{u_2} $. One can notice that even though we vary the fix values of $ \bm{\sigma}_{u_1} $ and $ \bm{\sigma}_{u_2} $ we see a little deviation of the results in terms of the DIC, LPML, and MSE values of the fitted integrated model. This follows that when the values of $ \bm{\sigma}_{u_1} $ and $ \bm{\sigma}_{u_2} $ are with in the range of (0, 2], then the integrated model is not affected by the fixed values of these parameters.  

\begin{table}[h]
	\centering
	\caption{DIC, LPML, and MSE of the integrated model for simulated data under different values of $ \bm{\sigma}_{u_1}^2 $ and $ \bm{\sigma}_{u_2}^2 $, censoring rate = 24\%}
	\label{table_sensitivity}
	\begin{tabular}{cccc}
		\hline
		\hline
		($ \bm{\sigma}_{u1}^2, \bm{\sigma}_{u2}^2 $) & DIC & LPML & MSE   \\
		\hline
		\hline
		(0.25, 0.25) & 305.64 & -290.84 & 0.0419 \\
		(0.50, 0.50) & 305.73 & -290.33 & 0.0417 \\
		(0.75, 0.75) & 306.20 & -289.04 & 0.0422 \\
		(1.00, 1.00) & 306.33 & -291.35 & 0.0417 \\
		(1.50, 1.50) & 307.32 & -289.34 & 0.0428 \\
		(2.00, 2.00) & 307.34 & -289.43 & 0.0423 \\
		\hline
		\hline
	\end{tabular}
\end{table}

\section{Circadian Genes from TCGA} \label{section_real_data}

In TCGA data, among available omics expressions, the DNA copy number changes are collected via SNP-arrays and array comparative genomic hybridization (aCGH) and for Breast cancer data, only the first kind is available via the \textsf{R} package \package{TCGA2STAT} \citep{ying2015tcga2stat}. TCGA provides the gene expression data in several different form and among them we have considered the one which is measured via RNA-Sequencing technology preprocessed using the first pipeline and normalized to get continuous measurements which is known as RPKM (Reads Per Kilobase Million). The original data is the version-stamped standardized data sets hosted and maintained by the Broad Institute GDAC Firehose. 

Our study focuses on the circadian genes and their effects on the patients' survival. We collected 10 such gene expressions (Table \ref{table_cicadian_genes}) with the corresponding observed survival components, the age and the gender of 68 Glioblastoma tumor samples and 364 Breast tumor samples.  

\begin{table}[h]
	\caption{Circadian Genes used for TCGA data analysis.} 
	\centering
	\begin{tabular}{l|l}
		\hline
		\hline
		Genes & Description \\
		\hline
		\hline
		CRY1     & belongs to the flavoproteins superfamily that exists in all \\
		& kingdoms of life and act as light-independent inhibitors of \\
		& CLOCK-BMAL1 components of the circadian clock \\
		CRY2     & belong to the flavoproteins superfamily that exists in all \\
		& kingdoms of life and act as light-independent inhibitors of \\
		& CLOCK-BMAL1 components of the circadian clock \\
		CSNK1E   & the protein encoded by this gene is a serine/threonine \\
		& protein kinase and a member of the casein kinase I \\
		& protein family, whose members have been implicated \\
		& in the control of cytoplasmic and nuclear processes, \\
		& including DNA replication and repair \\
		DEC1     & transcriptional repressor involved in the regulation of the \\
		& circadian rhythm by negatively regulating the \\
		& activity of the clock genes and clock-controlled genes. \\
		MT2      & is a member of the metallothionein family of genes. \\
		& Proteins encoded by this gene family are low \\
		& in molecular weight, are cysteine-rich, \\
		& lack aromatic residues, and bind divalent \\
		& heavy metal ions, altering the intracellular \\
		& concentration of heavy metals in the cell \\
		NPAS2    & a proetin coding gene and a transcriptional activator   \\
		& which forms a core component of the circadian clock  \\
		PER1     & encodes the period circadian protein homolog 1   \\
		         & protein in humans \\
		PER2     & a member of the Period family of genes \\
		& and is expressed in a circadian pattern \\
		& in the suprachiasmatic nucleus  \\
		PER3     & expressed in a circadian pattern in the   \\
		         & suprachiasmatic nucleus (SCN),  \\
		& the primary circadian pacemaker in the mammalian brain  \\
		TIMELESS & is notable for its role in Drosophila for encoding TIM,  \\
		& an essential protein that regulates circadian rhythm  \\
		\hline
		\hline
	\end{tabular}
	\label{table_cicadian_genes}
\end{table}

\subsection{Glioblastoma Cancer Data Analysis}  \label{section_GBM}

Glioblastoma, also known as glioblastoma multiforme or grade IV astrocytoma, is a fast-growing, aggressive type of central nervous system tumor that forms on the supportive tissue of the brain and it is the most common grade IV brain cancer. In 2018, more than 23,000 Americans were estimated to have been diagnosed and among them 16,000 were estimated to have died from brain and other nervous system cancers \citep{siegel2018cancer}. Glioblastoma accounts for about 15 percent of all brain tumors and occurs in adults between the ages of 45 to 70 years. Among the available data about 27\% are right censored. 

\begin{figure}[h]
	\centering
	\includegraphics[height = 5cm, width = 8cm]{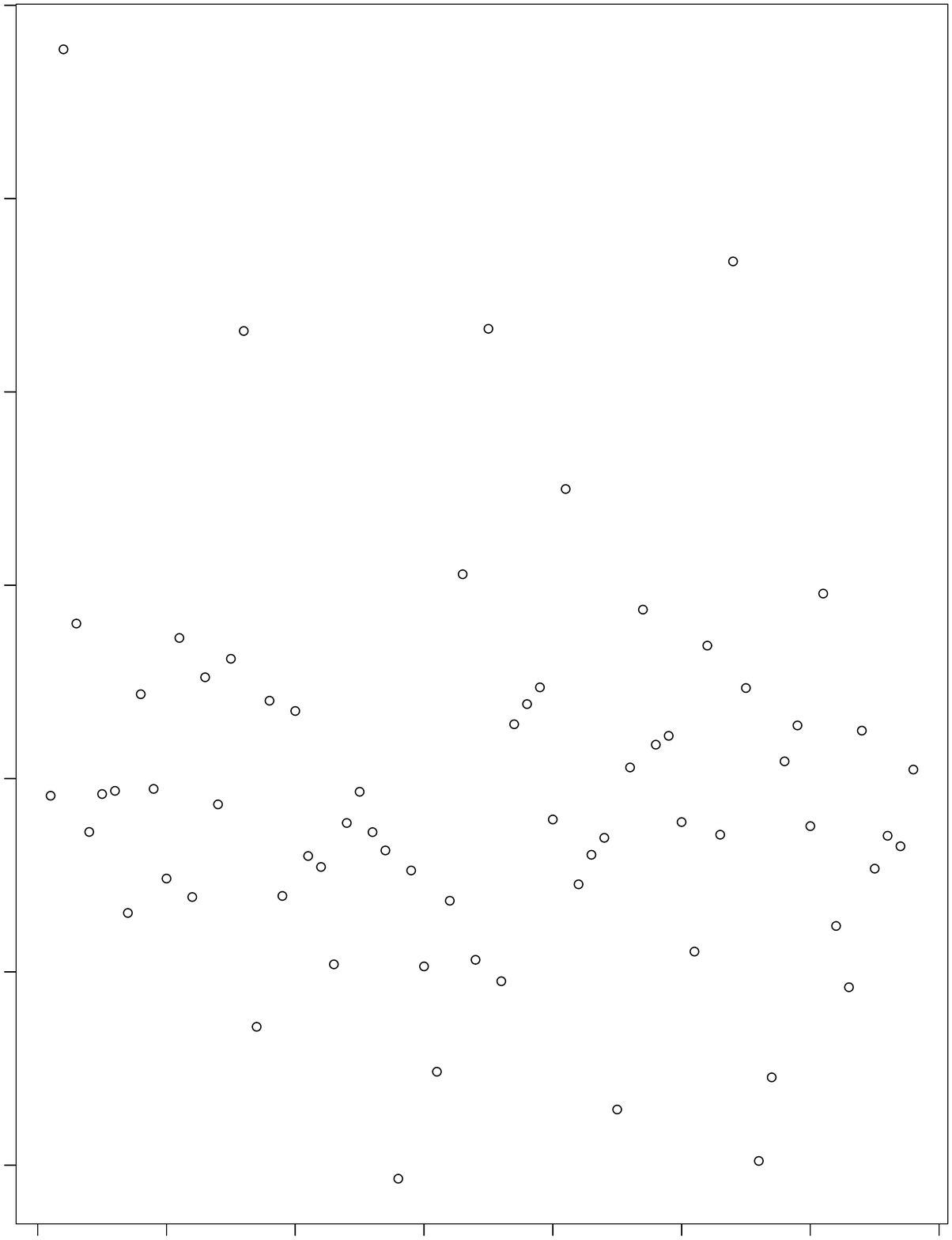}
	\centering
	\includegraphics[height = 5cm, width = 8cm]{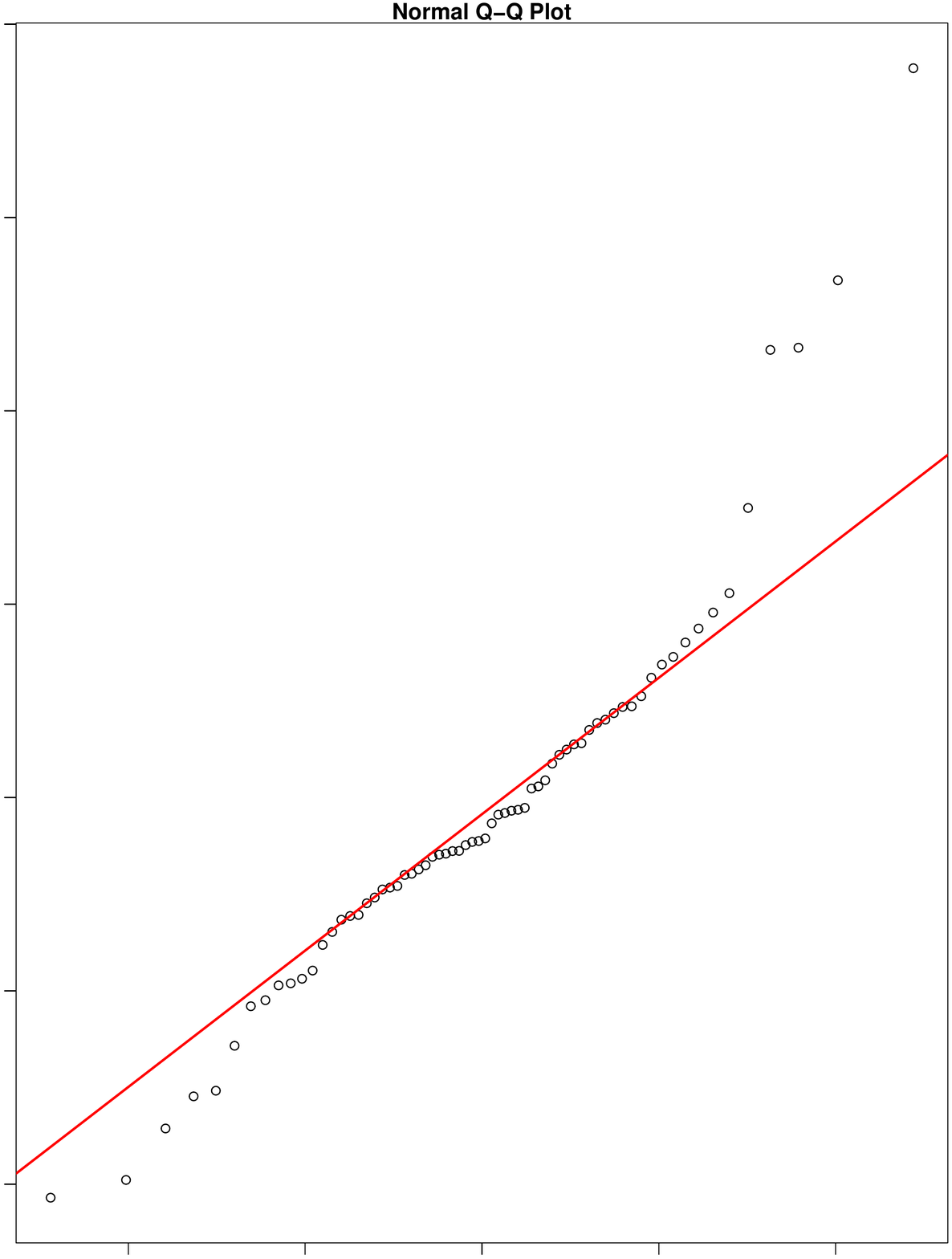}
	\caption{Top panel: Residual plot of the log normal regression of the survival time on the age of the individuals. Bottom Panel: QQ plot of those residual against the normal distribution.}
	\label{figure_residual}
\end{figure}

In addition, in this analysis, we consider the gender and the age of the patients as the external predictors on the survival time. As an exploratory analysis we fit a log normal AFT regression of the survival times of the individuals on their age. Figure \ref{figure_residual} displays the residuals and the QQ plot of those residuals. The residual plot shows that there is no clear pattern in the residuals. Furthermore, the QQ plot establishes that the log normal assumption on the residual distribution is adequate.

\begin{table}[h]
	\centering
	\caption{Goodness of fit for the integrated and nonIntegrated models in Glioblastoma data.}
	\begin{tabular}{lccc}
		\hline
		\hline
		Method & DIC & LPML & MSE \\
		\hline
		\hline
		integrated    & -244.93 & -175.58 & 0.230 \\
		nonIntegrated &  213.20 & -303.44 & 0.459 \\
		iBAG          & --      & --      & 0.392 \\
		\hline
		\hline
	\end{tabular}
	\label{table_GBM}
\end{table}

\begin{figure}[h]
	\centering
	\includegraphics[height = 5cm, width = 8cm]{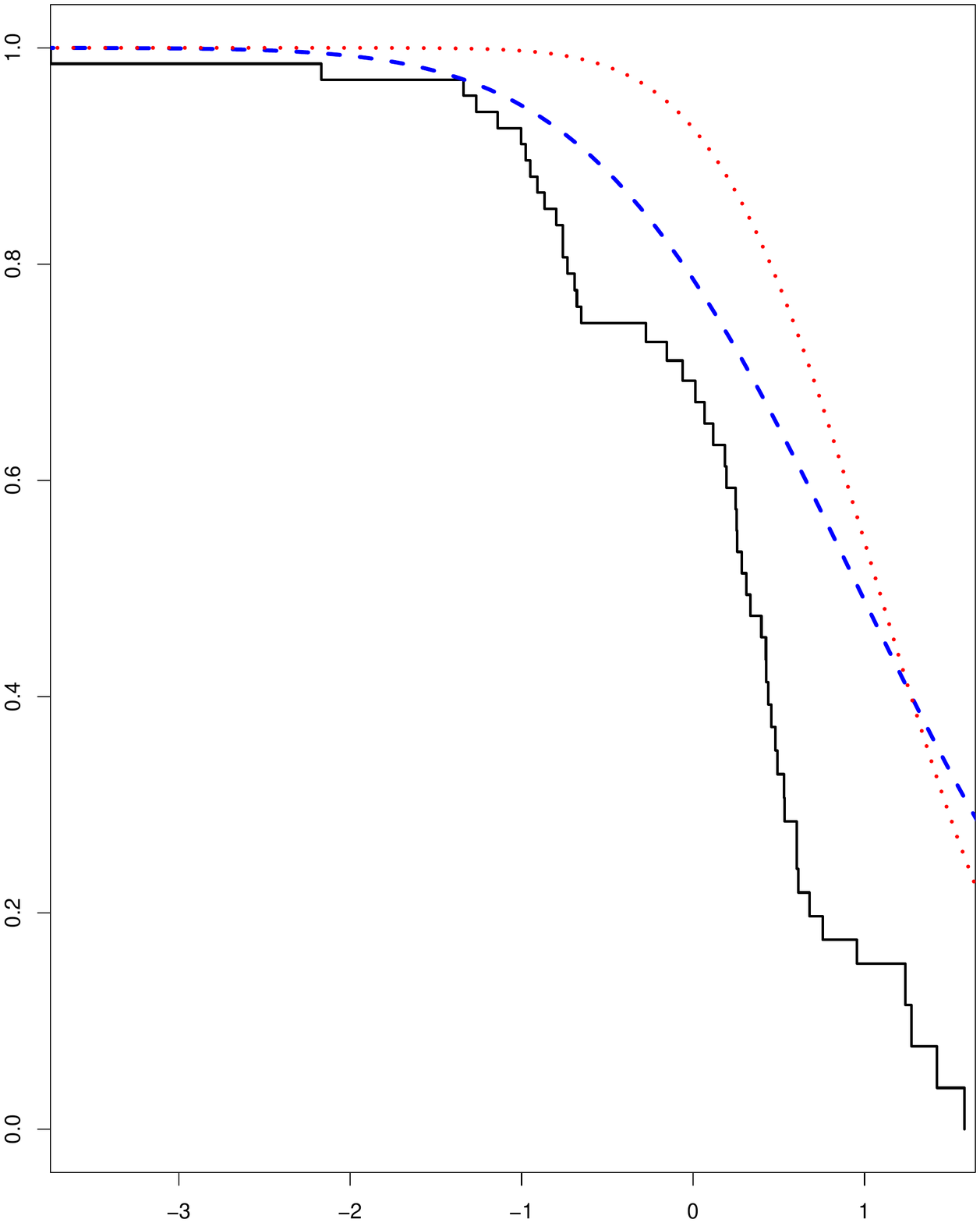}
	\centering
	\includegraphics[height = 5cm, width = 8cm]{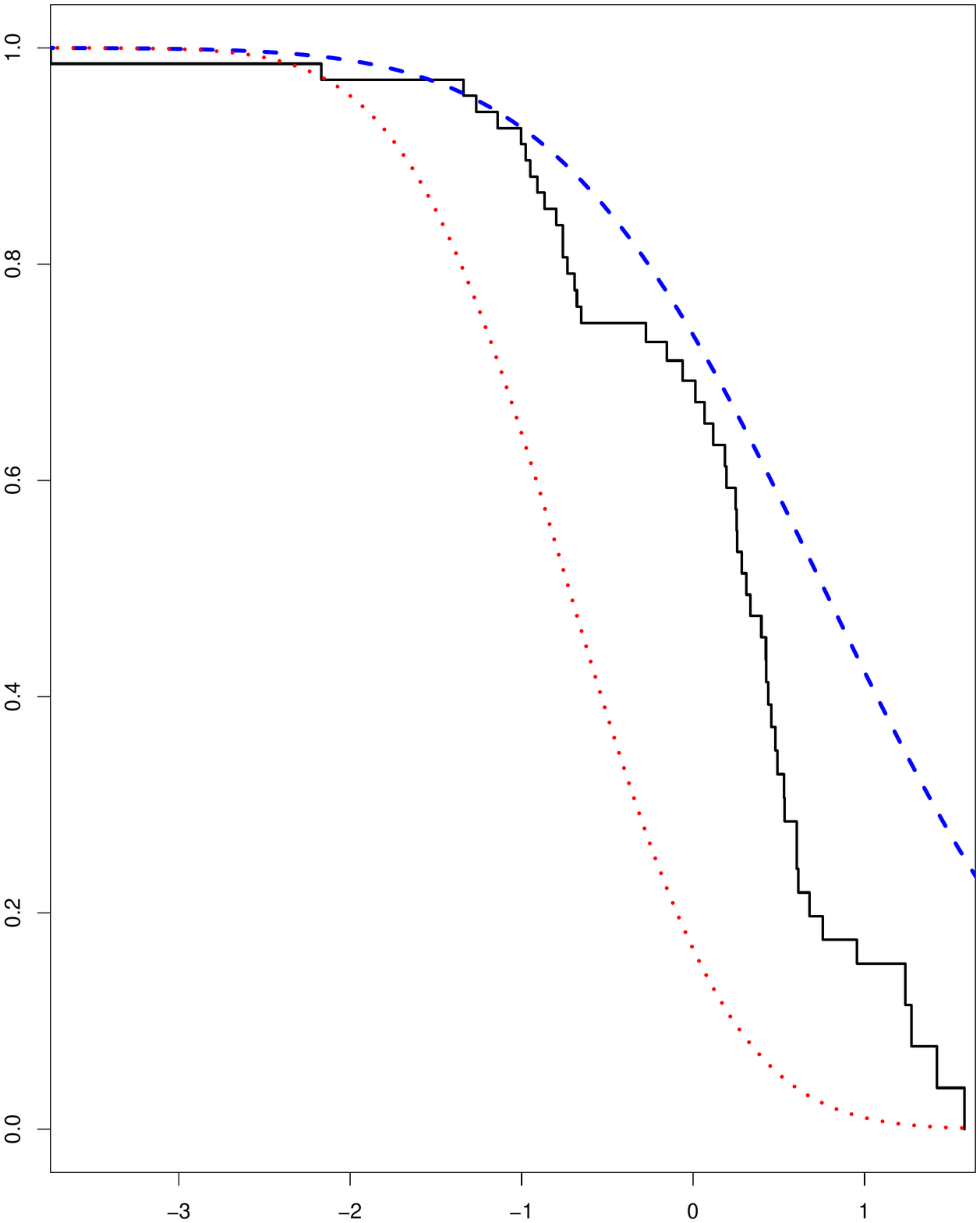}
	\caption{Survival functions for randomly selected two individuals for glioblastoma cancer dataset. Solid (black): the Kaplan-Meier plot, dashed (blue): integrated structural equation model, dotted (red): nonIntegrated model.}
	\label{figure_survial_gbm}
\end{figure}

We specify the following values for the prior distribution parameters for various parameters. For example, we set $ \bm{\beta}_{t0} = \bm{0} $, $ \Sigma_\beta $ is the unit variance covariance matrix, i.e., the diagonals are set to 1 and the off-diagonals are 0. We set, $ \alpha_{t0} = 0 $, $ \sigma_{\alpha_t}^2 = 1 $.  Similarly, the standard normal distribution is placed as the prior distributions of $ \alpha_{u_{1k}}, \alpha_{u_{2l}}, \phi_t, \phi_{u_{1k}}, $ and $ \phi_{u_{2l}}, $ $ k = 1, \ldots, q_1, $ $ l = 1, \ldots, q_2 $. For our experiment $ q_1 = q_2 = 10 $. 

We provide the goodness of fit results in Table \ref{table_GBM} and we note that the results of DIC and LPML suggests the superior performance of the proposed integrated approach compared to the traditional one. For example, when the SEM is fitted to the data the DIC is -244.93 which is lower than the DIC, 213.20, when the nonIntegrated model is fitted. Furthermore, in Figure \ref{figure_survial_gbm}, for two randomly selected individuals, we depict the survival probabilities computed using the two methods on the Kaplan-Meier plot. 

\subsubsection{Comparison with the Existing Method}

In this section we provide a brief study to find the performance of our proposed integrated structural equation model in comparison to the existing iBAG method \citep{wang2012ibag}. The experiment is carried out on the Glioblastoma dataset. We note that, iBAG method is primarily developed to assess the individual gene effect on the clinical outcome while considering the underlying relationship between the different high dimensional omics data platforms such as methylation and mRNA expressions. To this end this method employs a high dimensional Bayesian variable selection in the the fitting of the model. In contrast, in this article, the proposed structural equation method is examined only for circadian genes which are responsible for exhibiting time dependent behavior across 24 hours of each day, that is, we are interested in explaining the relationship between a particular trait and the survival of the cancer patients. Since feature selection is not the primary interest of our study a direct comparison is beyond the scope of this article. 

Nevertheless, in this example, we present a comparative study of both the methods. The computation for the iBAG method is carried out using the code given in \cite{wang2012ibag}. When fitting iBAG to the Glioblastoma dataset considered here, we replace the methylation expressions and the mRNA expressions with the copy number variation and the RNQSeq data respectively. The available program does not provide DIC and LPML for the fitted model. Hence the MSE is computed on the uncensored time points and is given in Table \ref{table_GBM}. One can note that the MSE due to our proposed integrated method is 0.230 and the same for iBAG method is 0.392. This concludes that the propose SEM method remains superior in terms of the prediction performance. 

\subsection{Breast Cancer Data Analysis}

Breast cancer is one of the most common cancers with a massive number of cases reported. For instance,  in 2018, more than 268,000 Americans were estimated to have been diagnosed and 41,000 were estimated to have died from breast cancer related tumors \citep{siegel2018cancer}. This heterogeneous disease is categorized into three groups such as the oestrogen receptor group, the HER2 amplified group, and the triple negative breast cancers or the basal like breast cancers \citep{cancer2012comprehensive}. Among them we consider the information of 364 breast tumor samples with their survival data from TCGA. We observe that at least 82\% data are right censored. In the analysis we consider the age variable as a covariate effect on the survival time. 

\begin{table}[h]
	\centering
	\caption{Goodness of fit for the integrated and nonIntegrated models in Breast cancer data.}
	\begin{tabular}{lcc}
		\hline
		\hline
		Method & DIC & LPML  \\
		\hline
		\hline
		integrated    & 1077.38 & -277.24  \\
		nonIntegrated & 1104.87 & -384.10  \\
		\hline
		\hline
	\end{tabular}
	\label{table_BRCA}
\end{table}

\begin{figure}[h]
		\centering
		\includegraphics[height = 5cm, width = 8cm]{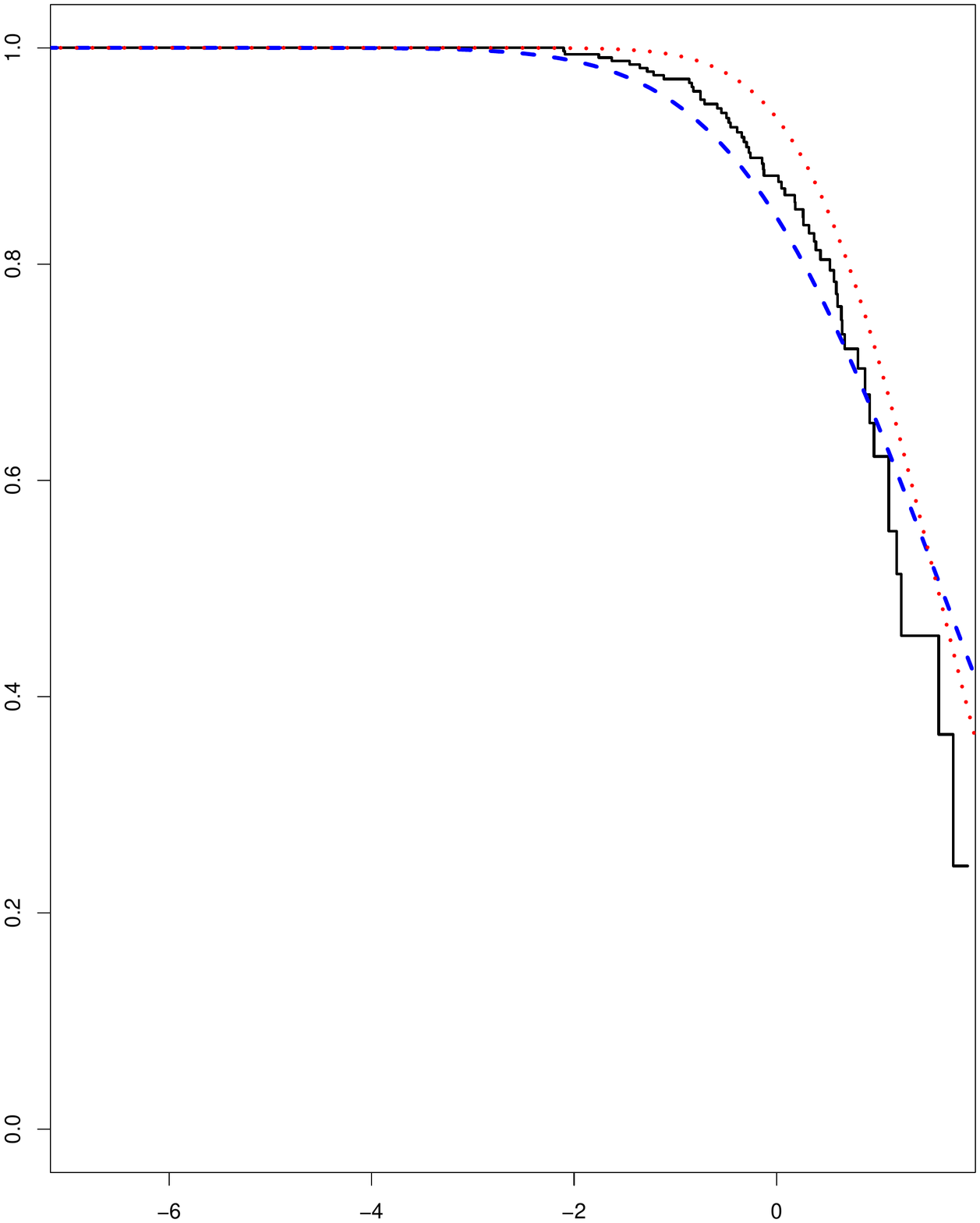}
		\includegraphics[height = 5cm, width = 8cm]{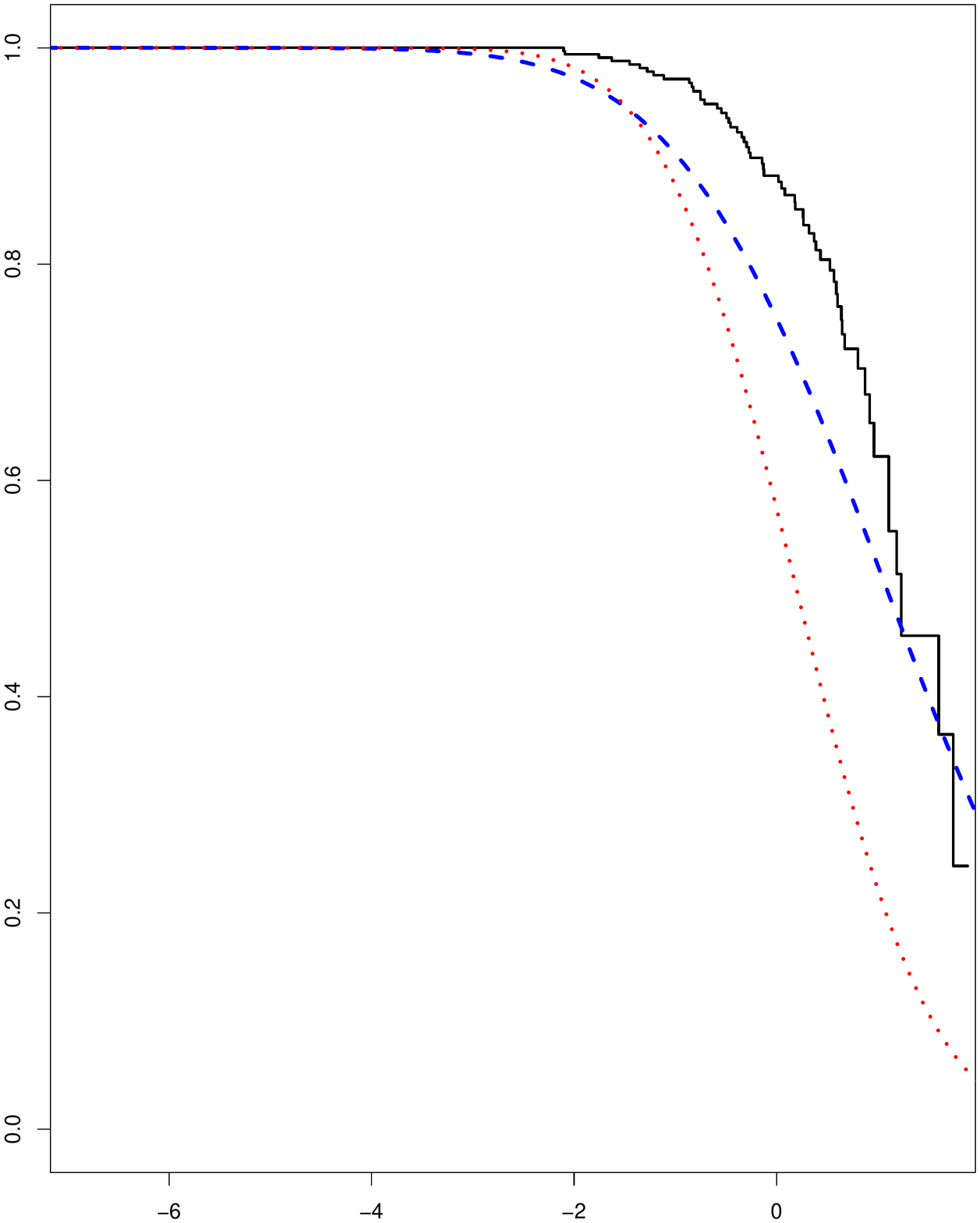}
	\caption{Survival functions for randomly selected two individuals for breast cancer dataset. Solid (black): the Kaplan-Meier plot, dashed (blue): integrated structural equation model, dotted (red): nonIntegrated model.}
	\label{figure_survial_brca}
\end{figure}

We present the goodness of fit results in Table \ref{table_BRCA} and we notice that the DIC due to our proposed method is 1077.38 which is less than the DIC 1104.87 due to the nonIntegrated model. This indicates that the proposed integrated model provides a better fit to the Breast cancer data. This is also confirmed by the LPML numbers obtained by fitting the different models to data. In Figure \ref{figure_survial_brca} for two randomly selected individuals, we depict the survival probabilities computed using the two methods on the Kaplan-Meier plot.

\section{Conclusion} \label{section_conclusion} 

In this article we have proposed a simple Bayesian structural equation modeling technique to integrate the information from different omics platform. We have shown that the proposed SEM technique provides improved survival prediction and better fits to the data compared to the traditional approach. Our focus in this article is concentrated on circadian genes only. Toward this end the sole intention of the proposed method is to capture the biological system in order to predict the patient survival when the circadian genes are of the interest. 

Nonetheless when a large number of gene expressions is under consideration and we have only limited number of patient samples then a sophisticated variable selection method needs to be implemented which will also have the ability to detect the effect of a single gene on the clinical outcome.

In a very general setup, we can allow a latent variable for each gene and use appropriate priors to borrow strength. This will be an over parameterized model with huge number of random effects and due to their correlations the computation will be extremely slow and expensive. The remedy is to categorize (cluster or group) the genes according to their functions and use a latent variable corresponding to each of these categories. In our applications, we are working only with circadian genes which can be treated as a single category and hence we have specified a single latent variable corresponding to it. Extension to multi-category models will be done in future research using clustered models.

We have specified a noninformative prior on $ \sigma_t^2 $. It is worth to mention that an Inverse Gamma prior would also maintain the conjugacy. However, our study shows that, imposing a suitable prior on all other variance parameters results in similar superior performance of the proposed structural equation based integrated modeling which is evident from the analysis given in the supplementary material. Hence, choice of appropriate priors for those parameters is kept for future studies. 

The two platforms we have considered here are RNAseq and CNV. In these regressions we separately regress the corresponding expressions on two separate latent variables for each gene. Hence we have assumed that those regressions are conditionally independent from each others. If a particular application violates this assumption caution should be exercised. 

We assume our model specification to be fully parametric. As a starting approach, the Log Normal model is assumed here. A Weibull model or a Gamma model is also possible to fit. However, all of these distributions have similar tail property. Moreover, we examine the residual plots of the Log Normal models (included in Section \ref{section_GBM} for the age variable and in the Supplementary Material for few genes) which are satisfactory for a Log Normal assumption. Nevertheless, one possible extension, as indicated by \cite{wong2018efficient}, is to consider nonparametric models which is due for the future research. The theoretical properties are also of future interests. 

The latent variables which are key components of the proposed model are platform specific, that is, each platform expression is regulated by a single latent variable which is sufficient for circadian oscillation characteristics. Using this and using the Log Normal AFT model we have developed the structural equation model to predict the clinical outcome survival. The Log Normal AFT model has been shown adequately fitted to the TCGA data considered here. In our examples we showed that the proposed model outperformed independent models. However, one must be aware that if any or some of the assumptions are not satisfied then the model should be tuned accordingly.

\section*{Supplementary Material}
The code for this paper is available at https://github.com/arnabkrmaity/sem\_mcmc/blob/master/mcmc.

\section*{Acknowledgments}

We are grateful to the editor, the associate editor Inanc Birol, and the three anonymous referees whose valuable comments have considerably improved this article. 

\vspace*{-12pt}

\section*{Funding}

The research reported in this paper has been supported by grants from the National Cancer Institute (R01-CA194391) and National Science Foundation (CCF-1934904).

\vspace*{-12pt}

\bibliographystyle{natbib}
\bibliography{sem_reference}

\clearpage

\end{document}